\documentclass[12pt]{article}

\usepackage[a4paper,margin=2.25cm]{geometry}
\usepackage{amsmath,amssymb}
\usepackage{graphicx}
\usepackage{microtype}
\usepackage{placeins}
\usepackage[numbers,sort&compress]{natbib}
\usepackage{xcolor}
\usepackage[colorlinks=true,allcolors=blue!55!black]{hyperref}

\newcommand{\softmax}{\operatorname{softmax}}

\title{\bfseries Temperature-driven inversion and nonlinear dynamics in ChatGPT-like AIs}
\author{Neil F. Johnson\textsuperscript{*},\quad Frank Yingjie Huo,\quad Bella Li\\[3pt]
\normalsize Department of Physics, The George Washington University\\
\normalsize Washington, DC 20052, USA\\
\normalsize \textsuperscript{*}Correspondence: \texttt{neiljohnson@gwu.edu}}
\date{}

\begin{document}
\maketitle

\begin{abstract}
Increasing the temperature of an ordinary many-state system increases access to a wider range of states and hence increases its entropy. We find the opposite in ChatGPT-like AIs, even though raising the decoder temperature \(T_d\) likewise increases access to a wider range of states (next-token choices). Across 12,000 continuations from 11 AIs, autoregressive feedback drives the long-time output population through an entropy maximum and into population inversion. The transition features frozen states, cycles, intermittency and noise-induced ordering.
We present evidence of a hidden coordinate $x_n$ that acts as the state variable of an effective nonlinear map. Its trajectory average strongly predicts output repetition in separate test trajectories.
ChatGPT-like AIs therefore behave not as `stochastic parrots', but as a new class of controllable nonlinear physical systems whose internal dynamics can be measured and perturbed.
\end{abstract}

ChatGPT-like AIs are doing increasingly consequential work in medicine and health, finance, business, law, science and defence~\citep{singhal2023,lopezlira2026,brynjolfsson2025,katz2024,wang2023discovery,nato2024}.  Yet all can produce fluent but undesirable output that can negatively affect diagnoses, livelihoods, legal rights, scientific conclusions and security decisions.  Safe use requires more than performance benchmarks: it demands a reproducible account of how these systems generate and change their output~\citep{bommasani2021}.

Existing studies chart output failures and architectural pathologies through benchmarks, theoretical analysis and model-specific diagnostics.  These include text degeneration and repetition, hallucination, loss of fidelity to earlier input, representation drift, and rank or depth collapse~\citep{holtzman2020,fu2021,guo2026,chen2025rank,sun2025depth,dong2021,wei2025,bang2025,alansari2025}.  Mechanistic-interpretability methods resolve individual internal features and circuits through controlled probes, sparse-autoencoder feature dictionaries, circuit tracing and attention-head visualization~\citep{nanda2023grokking,nanda2022grokking,nanda2023walkthrough,cunningham2024,ameisen2025,luger2026}.  The Jacobian Lens adds a layer-resolved readout, using averaged downstream Jacobians to expose interpretable hidden directions that can affect present and future output~\citep{gurnee2026}.  Yet exactly how these components combine to produce the observed output remains a mystery.  In the absence of a system-level account, one influential metaphor treats ChatGPT-like AIs as `stochastic parrots': systems that recombine statistical patterns in training text without grounded understanding~\citep{bender2021}.

At each next-token generation step $n$, a ChatGPT-like AI uses the prompt and all tokens generated before that step to assign a numerical score \(z_i\), called a logit, to every possible next token \(i\), and then selects one according to the resulting probabilities. The negative logit acts as an effective energy, \(E_i=-z_i\), so the decoder temperature \(T_d\) enters exactly as temperature does in a Boltzmann factor: token \(i\) receives a weight proportional to \(\exp(-E_i/T_d)\), normalized over all tokens~\citep{jaynes1957,hinton2015,huo2026atom,johnson2026simple}. For fixed logits, increasing \(T_d\) spreads probability across a wider range of possible next tokens, just as raising physical temperature spreads occupation across energy levels. With a large vocabulary providing many such effective levels, one would naturally expect increasing \(T_d\) to produce a monotonic increase in the entropy of the long-time output. As we show, this is not what happens.

Across 11 AIs, we show that decoder temperature $T_d$ acts as a controlled physical parameter that drives the long-time output population through an entropy maximum and into inversion, with the transition structured by frozen states, cycles, intermittency and noise-induced ordering.  Focusing on the large open-source model Llama-3.1-70B, we identify a hidden coordinate $x_n$ that predicts repetition and can be causally steered.  Together, these results show that the popular `stochastic parrot' metaphor is misleading: apparent randomness is instead organized by measurable transitions, memory and causal response.  

Our findings bring the massive societal challenge of understanding ChatGPT-like AIs into the established physics of controllable stochastic nonlinear systems whose internal dynamics can be measured and perturbed.
Nonlinear dynamics explains how mechanistic update rules generate fixed points, cycles, intermittency, bifurcations and chaos~\citep{strogatz2024}.  It has expanded from low-dimensional maps to high-dimensional network control~\citep{motter2012,motter2015} and brain dynamics~\citep{sporns2010}.  ChatGPT-like AIs --- autoregressive decoders that generate each next token from the prompt and all preceding tokens --- constitute a new class of iterative nonlinear systems with recomputed interactions and perturbable internal states.  Prior work reports nonlinear autoregressive formulations~\citep{krishnamurthy2026}, temperature-driven transitions and criticality~\citep{arnold2024,nakaishi2024,mikhaylovskiy2025,sun2025,ruan2026}, attractor cycles~\citep{wang2025}, chaos diagnostics~\citep{jaca2026} and causal repetition units~\citep{hiraoka2025}.  By contrast, our work explicitly unifies three previously separate levels: external control by the decoder temperature, collective output inversion and internal nonlinear dynamics along a measurable, perturbable hidden coordinate.

\section*{Temperature drives population inversion}

\begin{figure}[t]
\centering
\includegraphics[width=\linewidth]{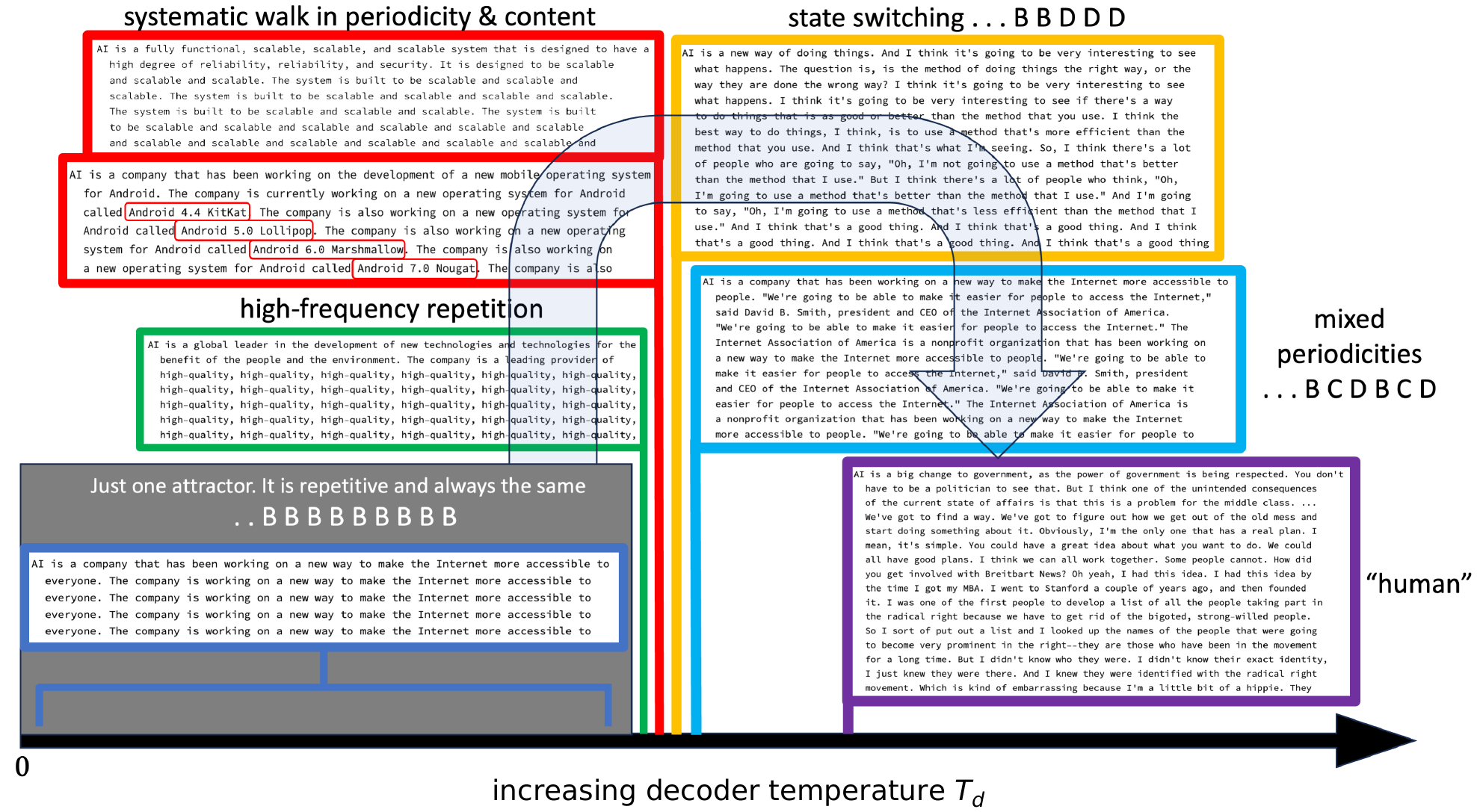}
\caption{\textbf{Illustrative GPT-2 output for the prompt `AI is \ldots'.}
As $T_d$ rises, generation moves from a single repetitive attractor-like pattern through additional periodicities and switching, and eventually to regularity and novelty coexisting in human-like output.  For convenience, we coarse-grain sentences to symbols. Figure~\ref{fig:spectroscopy} quantifies the resulting population inversion between coarse recurrent and complex/noisy macrostates.}
\label{fig:overview}
\end{figure}

Figure~\ref{fig:overview} illustrates how increasing decoder temperature $T_d$ changes GPT-2 output for the prompt `AI is \ldots'.  At low \(T_d\), a trajectory remains trapped in one repetitive basin over a range of temperatures.  Raising \(T_d\) opens other recurrent patterns and switching routes.  At still higher \(T_d\), recurrence and novelty coexist in prose with the balance of order and disorder characteristic of human language.  This suggests a transfer of the output-text population between a recurrent sector and a complex/noisy sector, where ‘population’ means the collection of complete texts generated over many runs.

The key nonlinear ingredient is autoregressive feedback: each sampled token is appended to the input for the next step, giving the AI an evolving memory of its own output.  If \(C_n\) is the prompt together with all tokens generated before step $n$, generation is the stochastic map \(C_n\mapsto C_{n+1}\): sample the next token from \(\softmax[z(C_n)/T_d]\), where \(z(C_n)\) is the vector of logits, append it to \(C_n\), and repeat.  Changing $T_d$ therefore changes both the current choice and the sequence of inputs subsequently explored, allowing the long-time output population to reorganize.  

\begin{figure}[t]
\centering
\includegraphics[width=\linewidth]{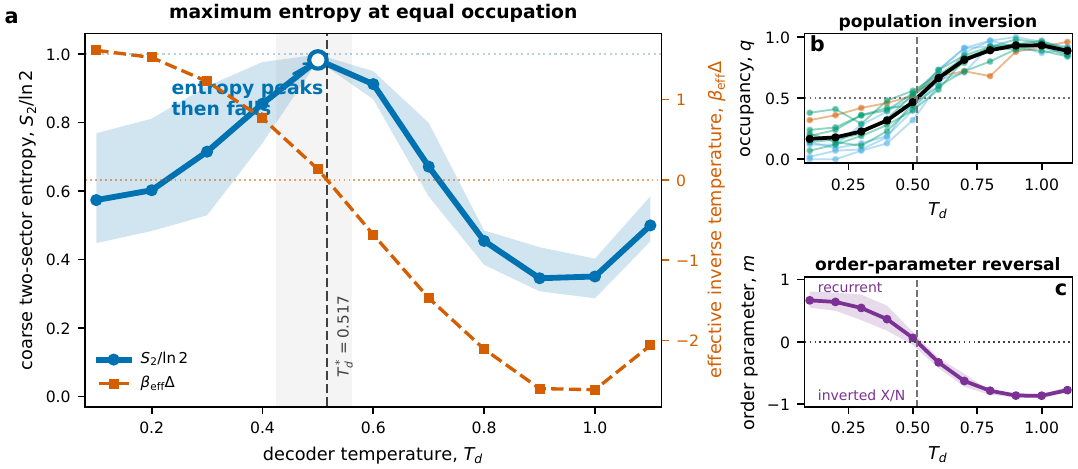}
\caption{\textbf{Maximum entropy marks ChatGPT-like AIs' output inversion.}
\textbf{a}, Normalized mixing entropy for the two output sectors.  Its maximum is near equal occupation.  The dashed orange curve is the occupancy-defined dimensionless parameter $\beta_{\mathrm{eff}}\Delta$, which changes sign at the mean crossing $T_d^*=0.517$; the grey band spans the 11 individual-AI crossings.
\textbf{b}, Occupancy of the complex/noisy sector for the 11 AIs (GPT-2, blue; Pythia, green; Llama, orange).  Black is the unweighted mean across AIs.
\textbf{c}, Reversal of the corresponding signed order parameter, completing the population-inversion signature.  Blue and purple shading in panels \textbf{a} and \textbf{c} spans the interquartile range across the 11 AIs.}
\label{fig:spectroscopy}
\end{figure}

To quantify this at the output level, we generated 12,000 continuations from four GPT-2 AIs, six Pythia AIs and Llama-3.1-70B using five fixed prompts~\citep{radford2019,biderman2023,llama3}.  Each continuation was classified by its sentence-level recurrence as frozen, sparse, periodic, intermittent, complex or noisy.  Classifying completed sentences in this way averages over fast token variation to expose slower recurrent output structures. Supplementary Note~2 gives the full construction.  AI-resolved crossings and family-level occupancy curves are reported in Supplementary Table~S2 and Supplementary Figure~S1. The complete 605-cell common-grid atlas is Supplementary Figure~S2, while Supplementary Table~S1 provides the complete evidence map.  For the population analysis, frozen, sparse, periodic and intermittent output effectively form a recurrent sector, while complex and noisy output effectively form a complex/noisy sector.  Let $q$ be the fraction in the complex/noisy sector.  The corresponding coarse two-sector variables are
\begin{equation}
 m=1-2q,\qquad
 S_2=-q\ln q-(1-q)\ln(1-q),\qquad
 \beta_{\mathrm{eff}}\Delta=\ln\!\frac{1-q}{q}.
 \label{eq:coarse}
\end{equation}
Here $m$ is the signed order parameter and $S_2$ is the entropy of the coarse two-sector population split.  $\Delta>0$ is the effective level gap between the two sectors, with the complex/noisy sector higher, while $\beta_{\mathrm{eff}}$ is the effective inverse temperature inferred from their relative occupancies.  Only the dimensionless product $\beta_{\mathrm{eff}}\Delta$ is determined by $q$.  Thus $\beta_{\mathrm{eff}}$ is an occupancy-defined effective quantity, not the inverse decoder temperature $1/T_d$.  Every AI crosses equal occupation.  The individual crossings lie between $T_d=0.425$ and 0.562, the unweighted mean across AIs crosses at 0.517, and the mean complex/noisy occupation reaches 0.932 at $T_d=1$ (Fig.~\ref{fig:spectroscopy}).  The entropy therefore rises to its maximum and then falls as the complex/noisy sector becomes overpopulated.

The robust transfer of the output population through equal occupation is the central empirical result; the entropy maximum and change of sign of $\beta_{\mathrm{eff}}\Delta$ are its statistical-mechanical signatures.  Beyond equal occupation, raising the positive decoder temperature reduces the coarse mixing entropy even though, at any specified next-token generation step $n$, it increases the entropy of the distribution for token $n+1$ when $C_n$ (the prompt and all tokens generated before step $n$) is held fixed.  In this two-sector mapping, the dimensionless inverse-temperature parameter $\beta_{\mathrm{eff}}\Delta$ changes sign and becomes negative.  The coarse output population therefore displays a bounded-spectrum inversion~\citep{purcell1951,ramsey1956,braun2013,abraham2017}: as $T_d$ increases, it passes from a recurrent-dominated sector through maximum mixing and becomes concentrated in the complex/noisy sector.  This collective reversal occurs despite the broadening of the token-$n+1$ distribution at each individual next-token generation step $n$.
\FloatBarrier

\section*{Structured routes through the transition}

\begin{figure}[!ht]
\centering
\includegraphics[width=0.78\linewidth]{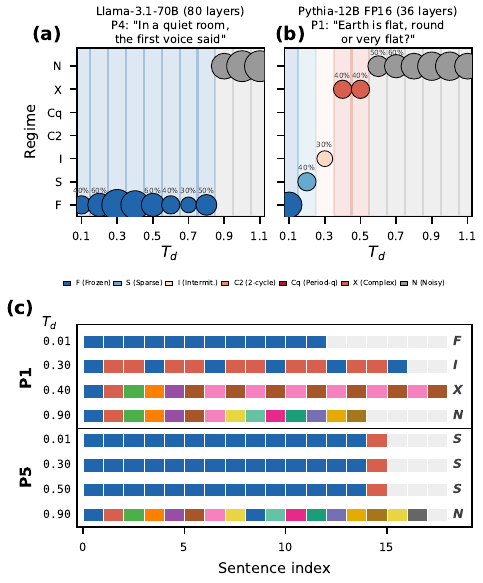}
\caption{\textbf{Different structured routes through the same population inversion.}
P1, P4 and P5 denote prompts 1, 4 and 5 in the fixed five-prompt set listed in Supplementary Note~2.
\textbf{a}, Frozen output is the most common regime for Llama-3.1-70B at low and intermediate temperatures, before stronger noise destroys that ordering.  At each $T_d$, the marker position and background colour identify the most common regime among ten runs; marker size gives the fraction of runs in that regime, with percentages shown when this fraction is below 65\%.
\textbf{b}, Pythia-12B instead passes through several recurrent regimes before noisy output becomes the most common; symbols and shading have the same meanings as in \textbf{a}.
\textbf{c}, Representative trajectories show frozen, sparse, intermittent, complex and noisy dynamics.  Each horizontal strip represents one run, and each coloured block represents one generated sentence.  Repeated colours within a strip identify sentences belonging to the same recurrent family; pale grey marks unused positions after generation ended.}
\label{fig:regimes}
\end{figure}

This population inversion (Fig.~\ref{fig:spectroscopy}a) is not a featureless crossover.  The six output classes reveal frozen or sparse recurrence, short cycles, intermittent switching, complex recurrence and high-diversity noisy motion.  Which regimes dominate depends on the AI and prompt, so the same inversion can be reached through different dynamical routes.

We use two prompt-conditioned ensembles to illustrate these structured routes (Fig.~\ref{fig:regimes}).  For Llama-3.1-70B prompt P4, frozen output is the most common regime from $T_d=0.1$ to 0.8 before noisy output becomes the most common at higher temperature.  Extending this comparison to near-greedy decoding, independent word $n$-gram statistics establish that the frozen interval is a noise-induced ordering window~\citep{horsthemke1984}, without sentence splitting or regime labels (Supplementary Fig.~S3).  Prompt-resolved occupancies show the heterogeneous routes hidden by averages over AIs (Supplementary Fig.~S4).  Pythia-12B prompt P1 instead passes through predominantly frozen, sparse and intermittent ensembles before complex and noisy output dominate.  Decoder temperature $T_d$ therefore reorganizes the recurrent structures sustained by feedback, rather than simply increasing disorder.
\FloatBarrier

\section*{Hidden coordinate predicts and steers output}

The output transition also has an internal dynamical coordinate $x_n$, measured immediately before token $n+1$ is selected at next-token generation step $n$.  For base Llama-3.1-70B under a fixed prompt, we generated trajectories of up to 300 tokens at 14 positive temperatures, using ten seeds to define the hidden direction and ten disjoint seeds to test the resulting coordinate; the full experimental configuration is listed in Supplementary Table~S3.  Repetition was measured directly as the fraction of repeated four-word sequences.  From training data only, we formed the hidden direction by joining the average normalized hidden states of the upper and lower quartiles of trajectories ranked by repetition.  We obtained $x_n$ by centring each later normalized hidden state on the training mean and projecting it onto this direction.  No test trajectory was used to define the direction, centre or thresholds.

\begin{figure}[t]
\centering
\includegraphics[width=\linewidth]{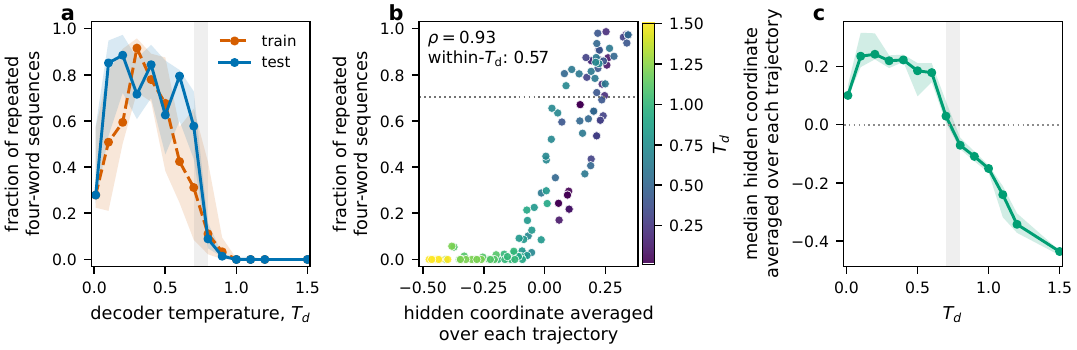}
\caption{\textbf{A hidden coordinate defined from training data predicts repetition in separate test trajectories.}
\textbf{a}, Repetition falls sharply between $T_d=0.7$ and 0.8 in separate training and test runs.  Lines show medians across ten runs at each temperature; shading spans the interquartile range.
\textbf{b}, Repetition in each of the 140 separate test trajectories versus the value of $x_n$ averaged over that trajectory; colour denotes decoder temperature.
\textbf{c}, The median trajectory-averaged value of $x_n$ reverses across the same sharp repetition change; shading spans the interquartile range across the ten test runs at each temperature.}
\label{fig:hidden}
\end{figure}

On 140 separate test trajectories, the value of $x_n$ averaged over each trajectory predicts repetition with rank correlation 0.933 (95\% interval, 0.914--0.948; Fig.~\ref{fig:hidden}); Supplementary Table~S4 gives the complete temperature sweep for the test trajectories.  The correlation remains 0.565 after subtracting the mean within each temperature, so the coordinate distinguishes trajectories generated at the same external temperature.  The median of these trajectory-averaged values also changes sign across the sharp loss of repetition between $T_d=0.7$ and 0.8.  Because the macroscopic population and hidden coordinate $x_n$ are constructed independently, their aligned change from recurrent to non-recurrent behaviour links the population inversion to the AI's internal dynamics.

A generic effective map follows by projecting the transformer operations onto the hidden direction (see Supplementary Note~6 for the details).  During each next-token generation step $n$, the attention calculation supplies learned token--token interactions through the softmax operation, a normalized exponential competition; layer normalization and the multilayer perceptron (MLP) then rescale and reshape the internal state.  Projected onto a direction separating two output sectors that compete near the transition, the attention update has an S-shaped response whose leading nonlinear saturation near balance is cubic.  Appending the sampled token changes the attention weights at next-token generation step $n+1$, converting the nonlinear calculation during step $n$ into a map from $x_n$ to $x_{n+1}$.  Slower internal variables appear as memory and sampling as noise.  This yields
\begin{equation}
 x_{n+1}\simeq a+b x_n-c x_n^3+\kappa x_{n-1}+\eta_n,
 \qquad c>0.
 \label{eq:nonlinear}
\end{equation}
Here $n$ indexes successive next-token generation steps and $x_n$ is the token-level hidden coordinate, oriented so that positive $x_n$ corresponds to greater repetition.  The constant $a$ biases the next step: $a>0$ favours repetition, whereas $a<0$ favours non-repetitive output.  The coefficient $b$ is the feedback gain for small displacements about the expansion point: $b>0$ tends to carry a displacement forward, whereas $b<0$ tends to reverse it and promote alternating motion.  Its magnitude, together with $\kappa$, controls whether a displacement decays or grows.  The cubic term with $c>0$ limits that growth, $\kappa$ represents short-term memory, and $\eta_n$ contains unresolved fluctuations.  The coefficients may change with decoder temperature.  Equation~\eqref{eq:nonlinear} therefore captures the leading bias, gain, saturation, memory and noise of the reduced dynamics.

We tested six candidate forms of this nonlinear map, fitting each to the training trajectories and testing them on entirely separate trajectories.  Maps using only the current coordinate, whether linear, polynomial or spline, perform similarly, whereas including $x_{n-1}$ improves prediction on the separate test trajectories at all 14 temperatures by 1.1--8.0\% (median 4.6\%) and reduces residual correlation.  Full diagnostics are shown in Supplementary Figure~S6.  The hidden coordinate $x_n$ therefore changes predictably and retains information from next-token generation step $n-1$: it is not merely a label assigned after the text has been generated.

\begin{figure}[t]
\centering
\includegraphics[width=\linewidth]{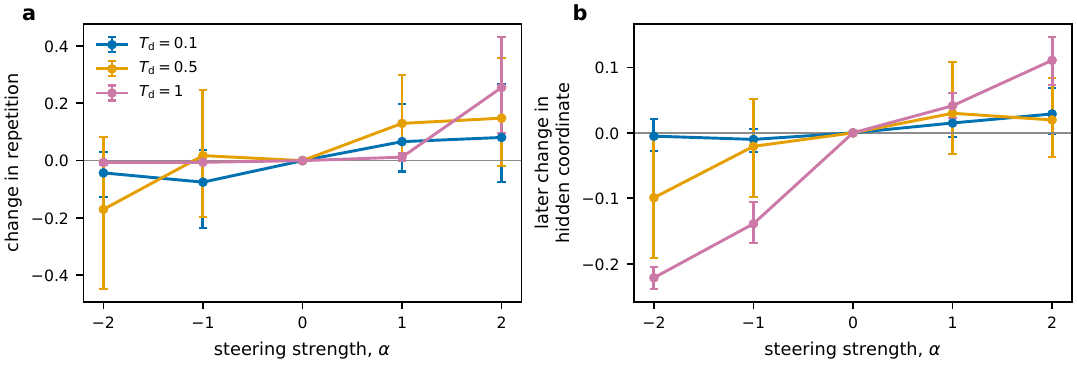}
\caption{\textbf{Steering the AI's hidden state changes repetition.}
\textbf{a}, Change in repetition as the hidden state is steered with different strengths along the hidden direction.
\textbf{b}, Resulting change in the hidden coordinate \(x_n\) at later next-token generation steps, measured immediately before the perturbation is applied at each step. Points show averages across ten separate test runs; error bars show 95\% bootstrap confidence intervals.}
\label{fig:causal}
\end{figure}

To verify that this hidden direction actually influences the output, we perturbed the AI along it at five steering strengths, $\alpha=-2,-1,0,1,2$, during text generation, repeating the experiment at three values of $T_d$.  The hidden coordinate $x_n$ is the scalar position along this direction at next-token generation step $n$.  The dimensionless parameter $\alpha$ sets the perturbation: its sign determines whether the perturbation is positive or negative along the hidden direction, while $|\alpha|$ sets its strength.  As a control, runs with $\alpha=0$ reproduced the corresponding unperturbed text sequences exactly.  Repetition and the later value of $x_n$ both increased monotonically on average: their mean slopes were 0.056 and 0.041 per unit $\alpha$, respectively, with aggregate estimates reported in Supplementary Table~S5.  At $T_d=1$, where the output population is strongly inverted, the largest positive steering strength, $\alpha=2$, increased repetition by 0.255 (95\% interval, 0.097--0.433), pushing trajectories back towards the recurrent sector (Fig.~\ref{fig:causal}).

\FloatBarrier

\begin{figure}[t]
\centering
\includegraphics[width=\linewidth]{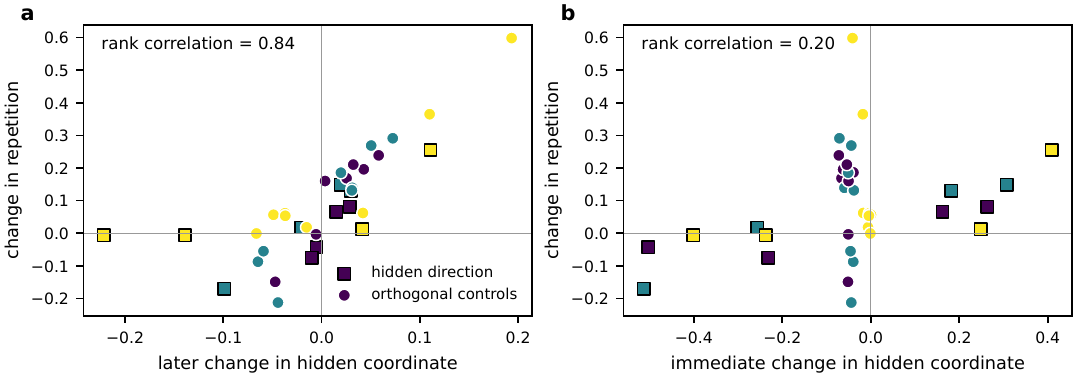}
\caption{\textbf{Behavioural effects align with the later change in the hidden coordinate $x_n$ across perturbations.}
\textbf{a}, Behavioural change closely tracks the later change in $x_n$ after autoregressive evolution.
\textbf{b}, It is only weakly related to the immediate change in $x_n$.  Each point represents one combination of perturbation direction, steering strength and decoder temperature, averaged over ten independent runs.  Squares show perturbations along the hidden direction and circles show the four matched orthogonal perturbations; colour indicates decoder temperature.}
\label{fig:reaction}
\end{figure}

Four matched orthogonal perturbations provide a broader demonstration that the different microscopic interventions are organized by the hidden coordinate $x_n$ (Supplementary Table~S6).  Across 36 combinations of perturbation direction, steering strength and decoder temperature, behavioural change correlates 0.842 with the later change in $x_n$ but only 0.199 with its immediate change (Fig.~\ref{fig:reaction}).  Interventions that produce similar later changes in $x_n$ produce similar behavioural effects, showing that the hidden coordinate organizes the evolving system.

\FloatBarrier

\section*{A laboratory for nonlinear physics}

We establish a unified physical picture of ChatGPT-like AIs' output process.  Increasing decoder temperature \(T_d\) broadens the distribution of possible next tokens at each next-token generation step \(n\), but autoregressive feedback reorganizes the long-time dynamics.  ChatGPT-like AIs are therefore unusual systems for physics: greater randomness in each next-token choice does not simply increase long-time disorder, but instead reorganizes the collective output.

Beyond the societal need to understand their output, ChatGPT-like AIs provide a new controlled laboratory for nonlinear physics.  Decoder temperature and initial conditions can be tuned, internal states can be observed, and perturbations can be applied as the output evolves.  This creates a direct experimental route from collective output transitions to the hidden dynamics that generate them.
\newpage

\section*{Methods}

\paragraph{Cross-AI survey.}
The survey used GPT-2 (124M, 355M, 774M and 1.5B parameters), Pythia (160M, 410M, 1.0B, 1.4B, 2.8B and 12B) and Llama-3.1-70B, with five fixed prompts and up to 300 generated tokens.  Conditions through Pythia-2.8B used 20 seeds and the two largest AIs used ten.  Within each continuation, sentence similarity was calculated from term-frequency--inverse-document-frequency (TF--IDF) vectors of character 3--5-grams and converted into a symbolic recurrence trajectory.  A fixed rule set classified frozen, sparse, periodic, intermittent, complex and noisy regimes.  The complex/noisy fraction was pooled across prompts within each AI and temperature; every AI received equal weight in cross-AI summaries.

\paragraph{Hidden-state experiment.}
The exact base Llama-3.1-70B checkpoint and tokenizer versions are recorded in Supplementary Table~S3.  Generation used 4-bit weights, bfloat16 computation, full-vocabulary sampling and deterministic algorithms.  Ten training seeds and ten disjoint test seeds were fixed before analysis.  The first 16 next-token generation steps were excluded as an initial settling period; hidden-state measurements began at step 17.  The hidden direction was the normalized difference between training-state averages from the upper and lower quartiles of trajectory repetition.  The coordinate $x_n$ was obtained by centring each later normalized hidden state on the training mean and projecting it onto this direction.  Bootstrap refits confirmed that this direction was stable.

\paragraph{Dynamics and interventions.}
We compared six candidate dynamical maps for predicting \(x_{n+1}\) from the current hidden coordinate $x_n$.  The candidates used linear, polynomial or spline dependence on $x_n$; extended maps also included $x_{n-1}$ or additional directions extracted from the training hidden states.  The hidden direction and four orthogonal control directions defined from training data were then applied after the initial settling period.  Each perturbed run was compared with its exact zero-strength control at the same temperature and seed.  Aggregate intervals resampled the ten test seeds.  Complete classifier rules, dynamical-map diagnostics and intervention tests are given in the Supplementary Information.

\paragraph{Statistics.}
Cross-AI summaries give each of the 11 AIs equal weight.  Hidden-coordinate correlations use 140 separate test trajectories---ten seeds at each of 14 positive temperatures---with 95\% intervals obtained by bootstrap resampling entire seed clusters.  Intervention effects are paired within the ten test seeds at each condition, and their 95\% intervals are percentile bootstrap intervals obtained by resampling seeds.  Wilcoxon diagnostics reported in the Supplementary Information are two-sided.  Exact sample sizes and definitions of independent runs are given with each figure and in Supplementary Notes~2--9.

\paragraph{Use of generative AI tools.}
OpenAI ChatGPT and Codex were used to assist with rephrasing the flow of the arguments, checking the consistency of the notation and proof-reading the manuscript and code.  

\section*{Data availability}

The manuscript package contains the archived hidden-state analysis, macroscopic survey and sensitivity results, full experimental configuration and integrity checks, and the numerical data used to produce the figures.
For reproducibility, Supplementary Note~10 and the accompanying file list record the archive's complete contents and its SHA-256 digital checksum.

\section*{Code availability}

The manuscript package is posted on Zenodo at \url{10.5281/zenodo.21752973}. 
It includes the executed Colab notebook, the same notebook as a Python script, and a script that rebuilds the order-parameter, hidden-state, memory and intervention-response figures.

\section*{Author contributions}
N.F.J. and F.Y.H. conceived and designed the study.  N.F.J. and B.L. ran the experiments and generated the data.  N.F.J., F.Y.H. and B.L. analysed the data, and N.F.J. prepared the figures.  N.F.J. wrote the initial draft and supervised the project.

\vspace{-1.25\baselineskip}
\section*{Competing interests}
N.F.J. is a co-founder of d-AI-ta Consulting LLC, which provides advice on AI deployment. The company was not involved in this work, and its activities are unrelated to the research reported here. F.Y.H. and B.L. declare no competing interests.

\bibliographystyle{unsrtnat}
\bibliography{references}

\end{document}